\documentclass{kapproc} 

\usepackage{procps} 

\usepackage[dvips]{graphicx}

\upperandlowercase

\kluwerbib 

\begin{document}

\articletitle[Dust penetrated arm classes]
{Dust penetrated arm classes: insight from rising and falling rotation curves}

\author{Marc S. Seigar\altaffilmark{1}, David Block\altaffilmark{2} 
and Ivanio Puerari\altaffilmark{3}} 
 
\affil{\altaffilmark{1}Joint Astronomy Centre, 660 N. A'ohoku Place, 
Hilo, HI 96720\\
\altaffilmark{2}School of Computational and Applied Mathematics, 
University of the Witwatersrand, P.O. Box 60, Wits, Gauteng 2050, South Africa\\
\altaffilmark{3}Instituto Nacional de Astrofisica, Optica y Electronica, Calle
Luis Enrique Erro 1, 72840 Tonantzintla, Puebla, Mexico}

\begin{abstract}
We present near-infrared K-band images of 15 galaxies. We have performed a
Fourier analysis on the spiral structure of these galaxies in order to 
determine their pitch angles and dust-penetrated arm classes. We have also
obtained rotation curve data for these galaxies and calculated their shear
rates. We show that there is a correlation between pitch angle and shear rate
and conclude that the main determinant of pitch angle is the
mass distribution within the galaxy. This correlation provides a physical basis
for the dust-penetrated classification scheme of Block \& Puerari (1999).
\end{abstract}


\section{Introduction}

The classification of galaxies, i.e. Hubble type, 
 has traditionally been inferred in the 
optical regime, where dust extinction
still has a large affect, and the light is dominated by the young
Population I stars. Infrared arrays offer opportunities for
deconvolving the Population I and Population II morphologies, because
in the K-band (2.2 $\mu$m), dust extinction is minimal, and the light
is dominated by old Population II stars. The extinction at this
wavelength is only 10\% of that in the V-band (Martin \& Whittet 1990).

Hubble
type is not correlated with the Population II morphology, as 
confirmed by the near-infrared studies of de Jong (1996) and Seigar \& James
(1998a, b). Also, it has been shown that near-infrared morphologies of spiral
galaxies are often vastly different from their optical morphologies (Block \&
Wainscoat 1991; Block et al. 1994a; Thornley 1996; Seigar \& James 1998a, b;
Seigar, Chorney \& James 2003). Often galaxies with flocculent spiral 
structure in the optical appear to have Grand-Design spiral structure in the 
near-infrared (Thornley 1996; Seigar et al. 2003). This suggests that the 
optical morphology bears little resemblance to underlying stellar mass 
distribution.

Burstein \& Rubin (1985) showed that spiral galaxies can have one of three
different types of rotation curve, rising, flat or falling. From these rotation
curve types they derived three principle types of mass distribution and found 
that Hubble types Sa and Sb were amongst all three types in approximately equal
amounts. This supports the idea that the optical morphology is not
correlated with the underlying mass distribution in spiral galaxies.

The disk of a spiral galaxy can be separated into
two distinct components: the {\em gas-dominated} Population I disk, and the 
{\em star-dominated} Population II disk. The former component contains
features of spiral structure (OB associations, HII regions, and cold 
interstellar HI gas). In contrast, the 
Population II disk contains the old stellar population highlighting the 
underlying
stellar mass distribution (Lin 1971). One might expect, even in the absence of
appreciable optical depths, for the two morphologies to be very different,
since the near-infrared light comes from mainly giant and supergiant stars
(Rix \& Rieke 1993; Frogel et al. 1996).

One therefore needs a near-infrared classification scheme, such as the 
dust-penetrated class (Block \& Puerari 1999; Block et al. 1999)
to describe the Population II disk, as well as
Hubble type to describe the Population I disk. The dynamic interplay between
the two components (via a feedback mechanism) is crucial, and has been 
studied extensively (Bertin \& Lin 1996). A central aspect here is the likely
coupling of the Population I disk with that of the Population II disk via a
feedback mechanism (Pfenniger et al. 1996).

The theoretical framework to explain the co-existence of completely different
morphologies within the same galaxy when it is studied optically and in the
near-infrared is described by Bertin \& Lin (1996). 
A global mode (Bertin et al. 1989a, b) is composed 
of spiral wavetrains propagating radially in opposite 
directions. Thus a feedback of wavetrains is 
required from the center. The return of wavetrains back to the corotation 
circle is guaranteed by refraction, either by the bulge or because the inner
disk is dynamically warmer. In the stellar disk, such a feedback can be 
interrupted by the Inner Lindblad Resonance (ILR), which is a location where
the stars meet the slower rotating density wave crests in resonance with their
epicyclic frequency (Mark 1971; Lynden-Bell \& Kalnajs 1972). In the gaseous
disk, the related resonant absorption is only partial, so that some feedback
is guaranteed. Once the above described wavecycle is set up, a self-excited
global mode can be generated.

The tightness of the arms in the modal theory comes from the mass distribution
and rate of shear. Galaxies with more mass concentration, i.e. higher overall
densities (including dark matter) and higher shear, should have more tightly
wound arms.
If the disk is very light (low $\sigma$ where $\sigma$ is the disk density) the
mode can be very tight, and one is in the domain of small epicycles.
If one increases the mass of the disk one finds a trend towards more open
structures, but soon one runs the risk of a disk that is too heavy and a bar
mode results.

The goal of this paper is to highlight the expected 
correlation between the shear rate
in spiral galaxies (as derived from their rotation curves) and the 
near-infared spiral arm pitch angle. The dust
penetrated morphology depends on the near-infared spiral arm pitch angle, and 
so, such a correlation would provide a physical basis  for the dust
penetrated classification scheme of Block \& Puerari (1999).

\section{Decomposition and identification of modes}

We have observed a sample of 15 galaxies in the near-infrared K-band 
(2.2 $\mu$m). These objects were taken from the study of Mathewson et al. 
(1992), who observed rotation curves for 1355 southern hemisphere spiral 
galaxies. Our sample includes galaxies with 
different rotation curve types (rising,
falling and flat) and span as wide a range of optical Hubble Types as possible.
The images were observed at the United Kingdom Infrared Telescope (UKIRT) using
the UKIRT Fast Track Imager (UFTI) between 1--4 August 2001 and 11--12 March
2002.

The 2-D Fast Fourier decomposition of all the near-infrared images in this
study, employed a program developed by I. Puerari (Schroeder et al. 1994).
Logarithmic spirals are assumed in the decomposition.

The amplitude of each Fourier component is given by:
\begin{equation}
A(m,p)=\frac{\Sigma^{I}_{i=1}\Sigma^{J}_{j=1}I_{ij}(\ln{r},\theta)\exp{-(i(m\theta+p\ln{r}))}}{\Sigma^{I}_{i=1}\Sigma^{J}_{j=1}I_{ij}(\ln{r},\theta)}
\end{equation}
where $r$ and $\theta$ are polar coordinates, $I(\ln{r},\theta)$ is the
intensity at position $(\ln{r},\theta)$, $m$ represents the number of arms
or modes, and $p$ is the variable associated with the pitch angle $P$, defined
by $\tan{P}=-\frac{m}{p_{max}}$.

Our Fourier spectra corroborate earlier observational indications
(Block et al . 1994a, 1999; Block \& Puerari 1999) that there is indeed a 
ubiquity of $m$=1 and $m$=2 modes. Block \& Puerari 
(1999) proposed three 
principle archetypes for the evolved stellar disk of such galaxies. The first
of these, designated dust-penetrated class $\alpha$, covers the pitch angle
range $4^{\circ}<P<15^{\circ}$, the second, designated $\beta$, covers 
$18^{\circ}<P<30^{\circ}$ and the third, designated $\gamma$ covers 
$36^{\circ}<P<76^{\circ}$.

Those lopsided galaxies (where $m$=1 is a dominant mode) are designated 
L$\alpha$, L$\beta$ and L$\gamma$ according to the dust penetrated pitch angle.
Evensided galaxies (where $m$=2 is the dominant Fourier mode) are classified
into classes E$\alpha$, E$\beta$ and E$\gamma$. Higher
order harmonics are classified
as H3 (for $m$=3) and H4 (for $m$=4).

\begin{figure}[ht]
\caption{Greyscale images of the galaxies for which the FFT analysis was performed. The overlaid contours represent the FFT fit to the spiral structure.}
\end{figure}

The range of radii over which the Fourier fits were applied are selected to 
exclude the bulge or bar (where there is no information about the arms) and
extend to the outer limits of the arms in our images. Pitch angles are then
determined from peaks in the Fourier spectra, as this is the most powerful
method to find periodicity in a distribution (Considere \& Athanassoula
1998; Garcia-Gomez \& Athanassoula 1993). The images were first deprojected
to face-on. Figure 1 shows the images of the spiral
galaxies observed for this project, overlaid with contours representing the
results of the Fourier analysis. The dust-penetrated arm classes and
pitch angles are listed in Table 1.

\section{Discussion}

Block et al. (1999) showed the first evidence that the pitch angle 
(and therefore
dust penetrated arm class) of a spiral galaxy depends upon the
shear rate as derived from rotation curves, consistent with the theoretical
predictions of the modal theory (Bertin et al. 1989a, b; Bertin \& Lin 1996;
Fuchs 1991, 2000). 
The work presented by Block et al.
(1999) consisted of just 4 galaxies. Here we present a further 15 galaxies,
all with measured rotation curves. Their shear rates are derived from their
rotation curves as follows
\begin{equation}
\frac{A}{\omega}=\frac{1}{2}\left(1-\frac{R}{V}\frac{dV}{dR}\right)
\end{equation}
where $A$ is the first Oort Constant, $\omega$ is the rotational velocity,
and $V$ is the velocity measured at radius $R$. The value $A/\omega$ gives the
shear rate.

\begin{table}[ht]
\caption{Results from the Fourier analysis and rotation curve analysis of 15 spiral galaxies. Column 1 shows the name of the galaxy; Column 2 shows the derived dust penetrated class; Column 3 shows the Hubble type; Column 4 shows the pitch angle of the K-band spiral arms and column 5 shows the derived shear rate.}
\begin{center}
\begin{tabular}{l|l|l|l|l}
Galaxy		& Dust-penetrated	& Hubble	& $P_K$		& $A/\omega$	\\
		& class			& type		&		&		\\
\hline
ESO 515 G3	& L$\gamma$		& SBc		& 47.8$\pm$0.7	& 0.27$\pm$0.02	\\
ESO 574 G33	& E$\gamma$		& SBbc		& 39.9$\pm$1.0	& 0.37$\pm$0.02	\\
ESO 576 G51	& E$\beta$		& SBbc		& 30.4$\pm$1.9	& 0.47$\pm$0.02	\\
ESO 583 G2	& E$\beta$		& SBbc		& 28.4$\pm$1.0	& 0.47$\pm$0.02	\\
ESO 583 G12	& L$\beta$		& SBc		& 17.7$\pm$2.0	& 0.54$\pm$0.02	\\
ESO 602 G25	& L$\beta$		& Sb		& 21.8$\pm$2.0	& 0.45$\pm$0.02	\\
ESO 606 G11	& H4$\beta$		& SBbc		& 25.2$\pm$2.4	& 0.50$\pm$0.02	\\
IC 1330		& E$\gamma$		& Sc		& 37.8$\pm$1.1	& 0.35$\pm$0.02	\\
NGC 2584	& E$\beta$		& SBbc		& 29.7$\pm$3.0	& 0.50$\pm$0.02	\\
NGC 2722	& E$\beta$		& Sbc		& 32.8$\pm$3.4	& 0.46$\pm$0.02	\\
NGC 3456	& E$\gamma$		& SBc		& 38.0$\pm$0.6	& 0.31$\pm$0.02	\\
NGC 7677	& E$\alpha$		& SABbc		& 17.0$\pm$0.8	& 0.66$\pm$0.02	\\
UGC 14		& L$\beta$		& Sc		& 20.9$\pm$1.3	& 0.53$\pm$0.02	\\
UGC 210		& L$\beta$		& Sb		& 16.0$\pm$0.8	& 0.55$\pm$0.02	\\
UGC 12383	& E$\alpha$		& SABb		& 15.0$\pm$1.6	& 0.66$\pm$0.02	\\
\end{tabular}
\end{center}
\end{table}

\begin{figure}[ht]
\caption{Pitch angle versus shear rate. Circles represent the
15 galaxies presented here. Squares are 3 galaxies presented in Block et al.
(1999).}
\end{figure}

\begin{figure}[ht]
\caption{Pitch angle versus rotation curve type. The symbols are the same as
Figure 2}
\end{figure}

Figure 2 shows a plot of the shear rate of spiral galaxies versus the 
spiral arm pitch angle. As well as showing a good correlation, it is also
interesting to note how the galaxies seem to fall into 3 distinct areas
on this plot, according to both their shear rates and pitch angles, and
possibly their mass distributions. Galaxies,
with high shear rates (rising rotation curves) 
and tightly wound spiral structure are found in the 
bottom right and are designated $\alpha$ in the dust penetrated class. Galaxies
with shear rates around 0.5 (flat)
and moderately wound spiral structure are in the
middle and are designated $\beta$. The top left contains those galaxies with
loosely wound structure and low shear rates (falling). 
These are designated $\gamma$.
Figure 3 shows a plot between rotation
curve type (Burstein \& Rubin 1985), i.e. rising, flat or falling, 
versus spiral arm pitch angle, also showing a good correlation.

The shape of a rotation curve is determined largely by the distrbution of 
luminous and dark mass contained in a spiral galaxy. The correlation found
between pitch angle and shear rate, therefore suggests that the main
factor determing the tightness of spiral structure, is in fact the central
mass concentration. Essentially, a declining rotation curve is indicative of
a large central bulge. The correlation between mass concentration and spiral
arm pitch angle has been suggested by many theoretical models
(e.g. Fuchs 1991, 2000; Bertin et al. 1989a, b; Bertin \& Lin 1996).

\begin{acknowledgments}
The United Kingdom Infrared Telescope (UKIRT) is operated by the 
Joint Astronomy Centre on behalf of the U.K. Particle Physics and 
Astronomy Research Council (PPARC). The authors would
like to thank David Block, Ken Freeman and the scientific organising
committee for
the opportunity to present this work at the Bars Congress 2004.

\end{acknowledgments}


\begin{chapthebibliography}{1}
\bibitem{bertin1}
Bertin G., Lin C.C., Lowe S.A., Thurstans R.P., 1989a, ApJ, 338, 78

\bibitem{bertin2}
Bertin G., Lin C.C., Lowe S.A., Thurstans R.P., 1989b, ApJ, 338, 104

\bibitem{bertin}
Bertin G., Lin C.C., 1996, {\em Spiral Structure in Galaxies: A density wave theory}, MIT Press, Cambridge, MA

\bibitem{bw}
Block D.L., Wainscoat R., 1991, Nature, 353, 48

\bibitem{block3}
Block D.L., et al., 1994a, A\&A, 288, 365

\bibitem{block4}
Block D.L., et al., 1994b, A\&A, 288, 383

\bibitem{block1}
Block D.L., Puerari I., 1999, A\&A, 342, 627

\bibitem{block2}
Block D.L., et al., 1999, Ap\&SS, 269-270, 5

\bibitem{burstein}
Burstein D., Rubin V.C., 1985, ApJ, 297, 423

\bibitem{Considere}
Considere S., Athanassoula E., 1988, A\&AS, 76, 365

\bibitem{deJong}
de Jong R.S., 1996, A\&A, 313, 45

\bibitem{frogel}
Frogel J.A., Quillen A.C., Pogge R.W., in {\em New Extragalactic Perspectives in the New South Africa}, eds D.L. Block, J.M. Greenberg, Kluwer, Dordrecht, p251

\bibitem{fuchs1}
Fuchs B., 1991, in {\em Dynamics of Disk Galaxies}, ed B. Sundelius, Chalmers University of Technology, p359

\bibitem{fuchs2}
Fuchs B., 2000, in {\em Galaxy Dynamics: From the Early Universe to the Present}, eds F. Combes \& G. Mamon, ASP Conf. Ser. Vol. 197, p53

\bibitem{garcia}
Garcia--Gomez C., Athanassoula E., 1993, A\&AS, 100, 431

\bibitem{lin}
Lin C.C., 1971, in {\em Highlights of Astronomy Vol. 2}, ed. C. de Jager, Reidel, Dordrecht, p88

\bibitem{lynden}
Lynden-Bell D., Kalnajs A.J., 1972, MNRAS, 157, 1

\bibitem{mark}
Mark J.W-K., 1971, Proc. Natl. Acad. Sci., 68, 2095

\bibitem{martin}
Martin P.G., Whittet D.G.B., 1990, ApJ 357, 113

\bibitem{mathewson}
Mathewson D.S., Ford V.L., Buchhorn M., 1992, ApJS, 81, 413

\bibitem{pfenniger}
Pfenniger D., Martinet L., Combes F., 1996, in {\em New Extragalactic Perspectives in the New South Africa} eds D.L. Block, J.M. Greenberg, Kluwer, Dordrecht, p291

\bibitem{rix}
Rix H.-W., Rieke M.J., 1993, ApJ, 481, 123

\bibitem{schroder}
Schroeder M.F.S., Pastoriza M.G., Kepler S.O., Puerari I., 1994, A\&AS, 108, 41

\bibitem{seigar1}
Seigar M.S., James P.A., 1998, MNRAS, 299, 672

\bibitem{seigar2}
Seigar M.S., James P.A., 1998, MNRAS, 299, 685

\bibitem{seigar3}
Seigar M.S., Chorney N.E., James P.A., 2003, MNRAS, 342, 1

\bibitem{thornley}
Thornley M.D., 1996, ApJ, 469, 45

\end{chapthebibliography}
\end{document}